# Layered Photonic Topological Insulators

Xiao-Dong Chen, and Jian-Wen Dong[*]

*School of Physics & State Key Laboratory of Optoelectronic Materials and Technologies, Sun Yat-sen University, Guangzhou 510275, China.*

[*]Corresponding author: dongjwen@mail.sysu.edu.cn

**The recent realization of photonic topological insulators has brought the discovery of fundamentally new states of light and revolutionary applications such as non-reciprocal devices for photonic diodes and robust waveguides for light routing. The spatially distinguished layer pseudospin has attracted attention in two-dimensional electronic materials. Here we report layered photonic topological insulators based on all-dielectric bilayer photonic crystal slabs. The introduction of layer pseudospin offers more dispersion engineering capability, leading to the layer-polarized and layer-mixed photonic topological insulators. Their phase transition is demonstrated with a model Hamiltonian by considering the nonzero interlayer coupling. Layer-direction locking behavior is found in layer-polarized photonic topological insulators. High transmission is preserved in the bilayer domain wall between two layer-mixed photonic topological insulators, even when a large defect is introduced. Layered photonic topological insulators not only offer a route towards the observation of richer nontrivial phases, but also open a way for device applications in integrated photonics and information processing by using the additional layer pseudospin.**

Recently, the realization of classical analogues of topological insulators in artificial crystals has been an emerging research area [1-8]. Particularly in 2005, Haldane and Raghu transferred the key feature of topological insulators to the realm of photonics and proposed the photonic analog of quantum Hall effect in photonic crystals (PCs) [9, 10]. After this milestone of work, photonic analog of topological insulators, i.e., photonic topological insulators (PTIs) have been theoretically proposed and experimentally demonstrated in different photonic systems [11-25]. Some traditional views on light propagation were overturned with the application of topological concepts. For example, broadband non-reciprocal propagation of light was realized [9-12], and reflection-free routing of light along sharply bent waveguides was demonstrated [16-18]. By borrowing the discrete degrees of freedom (e.g., spin and valley), the advent of PTIs has opened a path towards the discovery of fundamentally new states of light and revolutionary photonic devices.

On the other hand, the emergence of two-dimensional layered materials has provided a laboratory for exploring topologically nontrivial phases such as quantum Hall, spin Hall, and valley Hall phases [26-29]. Recently, in addition to the real electronic spin and valley pseudospin, the layer degree of freedom has been investigated as an additional pseudospin in bilayer systems [30]. A state of bilayer graphene or transition metal dichalcogenides localized to the upper or lower layer can be labelled with pseudospin up or down, respectively. The exploration of layer pseudospin and its interplay with other degrees of freedom has inspired numerous intriguing phenomena such as tunable gap bandwidth [31], spin-layer locking effects [32], magnetoelectric effects [33], and unconventional superconductivity [34].

In this work, we report layered PTIs by introducing the layer pseudospin into all-dielectric bilayer PC slabs. The introduction of additional layer pseudospin offers more dispersion engineering capability and a model Hamiltonian is developed to describe the interlayer interaction within the bilayer PC slabs. Layer-polarized and layer-mixed PTIs are realized, and their phase transition is demonstrated. Layer-direction locking behavior of edge states is found at the boundary between two layer-polarized PTIs. High transmission is achieved in the bilayer domain wall between two layer-mixed PTIs, even when a large defect is introduced. Layered PTIs open up a route towards the discovery of fundamentally novel states of light and applications such as layer-dependent light transmission.

**Bilayer photonic crystal slabs**

We consider the bilayer photonic crystal (PC) slab which consists of two layers of dielectric PC slabs, i.e., the upper layer1 and the lower layer2 [Fig. 1a]. They have the thickness of $h_1$ and $h_2$, and are separated by a homogeneous spacer ($\varepsilon = 1$) with the thickness of $h_s$. Each PC slab consists of a dielectric plate ($\varepsilon = 12$) with periodic air-holes with the lattice constants of $a_1$ and $a_2$ (outlined by white dash rhomboids). There are two equilateral triangular air-holes in each slab, i.e., A1 and B1 in layer1 while A2 and B2 in layer2. The lateral sizes of the triangular air-holes in layer1 are $s_{A1}$ and $s_{B1}$, while those in layer2 are $s_{A2}$ and $s_{B2}$. Transverse-electric like states of each slab form a gapless Dirac cone at the corner of Brillouin zone [Fig. 1b, and see details in Supplementary Section A]. When $h_s \gg h_1$, there is no interlayer coupling between two slabs and two Dirac cones are doubly degenerate [Fig. 1c]. When the interlayer coupling is turned on, the band dispersions of bilayer PC slabs can be engineered. In this work, we keep equal lattice constants of $a = a_1 = a_2$ and focus on the AA-stacking bilayer structure [See the stacking morphology and AB-stacking bilayer structure in Supplementary Section B]. When the thickness of spacer becomes comparative to those of the PC slabs (i.e., $h_s \sim h_1$), the electromagnetic fields of two slabs are coupled, leading to two oppositely shifted Dirac cones along the frequency axis [Fig. 1d]. To see band dispersion engineering, we consider the AA-stacking bilayer PC slab with $a = 460$ nm, $h_1 = h_2 = h_s = 220$ nm, and equal air-hole sizes of $s_{A1} = s_{B1} = s_{A2} = s_{B2} = 250$ nm [inset of Fig. 2a]. These structural parameters are tuned to bring the bilayer PC slabs working at the telecommunication frequency around 200 THz. Figure 2a shows its bulk band structure with $k_z = 0$, with the light cone shaded in grey. Under the light cone, two Dirac cones are separated along the frequency axis. Figure 2b illustrates the $H_z$ fields of four bulk states at the K point. Two degenerate mirror-symmetric states (① and ② patterns) appear at the lower frequency, while two degenerate mirror-asymmetric states (③ and ④ patterns) show up at the higher frequency. Due to different mirror representations, two cones cross with each other. It leads that a frequency nodal ring between the second and third bands encloses around the K point, serving as an ideal starting point to have PTIs.

**Layered photonic topological insulators**

The all-dielectric bilayer PC slab presented in Fig. 2a has the $D_{6h}$ symmetry which contains the mirror symmetry and the inversion symmetry. Layered PTIs can be obtained by applying symmetry breaking perturbations to this bilayer PC slab. We first discuss the structure where the mirror or inversion symmetry is broken individually, and then in next section consider the general case where mirror and inversion symmetries are broken simultaneously.

As shown in the inset of Fig. 2c, we first consider the bilayer PC slab with $s_{A1} = s_{B2} = 130$ nm and $s_{B1} = s_{A2} = 328.8$ nm. For this structure, the inversion symmetry is preserved while the mirror symmetry is broken. The broken mirror symmetry leads to the repulsion between the second and third bands, and a resultant directional band gap is obtained between 185.2 and 206.2 THz (i.e., between 1620 and 1455 nm) [Fig. 2c]. Below the band gap, there are two degenerate states at the K point. These two states have fields localizing at different layers [Fig. 2d]. One state has $H_z$ fields localizing at layer1 and power flux rotating anticlockwise (see in ① pattern), while the other state has $H_z$ fields localizing at layer2 and power flux rotating clockwise (see in ② pattern). Hence, the two-fold degeneracy is between the pseudospin-up anticlockwise state and pseudospin-down clockwise state when the *layer* is introduced as a pseudospin [3D schematics in Fig. 2d]. It is the photonic analogue of spin-orbit coupling, and we denote this resultant pseudospin-Hall PTI as the layer-polarized PTI.

We next consider the bilayer PC slab with $s_{A1} = s_{A2} = 130$ nm and $s_{B1} = s_{B2} = 328.8$ nm in which the mirror symmetry is preserved while the inversion symmetry is broken [inset of Fig. 2e]. Figure 2e shows its bulk band structure in which a directional band gap is also found. Below the band gap, both two bulk states at the K point have their $H_z$ fields localizing at B1 and B2 air-holes [Fig. 2f]. Induced by the nonzero interlayer coupling, fields are mixed between two layers. In addition, indicating by the same anticlockwise power fluxes [grey arrows], these two states belong to the same orbital states [3D schematics in Fig. 2f]. Combining these properties, we denote this kind of bilayer PC slab as the layer-mixed PTI.

**Phase transition and effective Hamiltonian**

The layer-polarized PTI presented in Fig. 2c and the layer-mixed PTI shown in Fig. 2e are topologically distinct. Here, we simultaneously apply both mirror and inversion symmetry breaking perturbations to observe the phase transition. To do this, we consider the bilayer PC slabs with varied $s_{A1}$ and $s_{A2}$. The amplitude of mirror (inversion) symmetry breaking is given by $\Delta = s_{A1}^2 - s_{A2}^2$

($\Sigma = s_{A1}^2 + s_{A2}^2 - 125000\,\text{nm}^2$). In addition, we keep the same air-holes sizes at each slab, i.e., $s_{A1}^2 + s_{B1}^2 = s_{A2}^2 + s_{B2}^2 = 125000\,\text{nm}^2$. This constant filling ratio condition ensures the dispersions of perturbed bilayer PC slabs do not move much in frequency, and keeps the mid-gap frequency unchanged. A phase diagram is shown in Fig. 3a in which the horizontal and longitudinal axis are respectively given by $\Delta$ and $\Sigma$. The bilayer PC slab discussed in Fig. 2a locates at the origin of phase diagram, and it has two pairs of gapless linear dispersions near the K point [Fig. 3b4]. Along the horizontal axis, a directional band gap is immediately obtained for a nonzero $\Delta$ [Fig. 3b1]. The larger the absolute value of $\Delta$, the larger the band gap, e.g., the one presented in Fig. 2c. This kind of band gap belongs to the layer-polarized PTI and it is shaded in cyan in the phase diagram. On the contrary, along the longitudinal axis, the frequency nodal ring still exists for bilayer PC slabs with small $\Sigma$ [Fig. 3b5]. Such nodal ring will be split out by further increasing the absolute value of $\Sigma$. Lastly, a resultant band gap is obtained [Fig. 3b6]. By further increasing the absolute value of $\Sigma$, the bandwidth of band gap can be enlarged. One example has been illustrated in Fig. 2e in which the layer-mixed PTI is found [shaded in green]. The phase transition between these two kinds of band gaps can be found by tuning the amplitudes of broken mirror and inversion symmetry. For example, starting at the inversion symmetry breaking dominated phase [Fig. 2b6], we gradually decrease the amplitude of broken inversion symmetry while increase the amplitude of broken mirror symmetry. The second and third states meet each other [Fig. 2b3], then move apart and evolve into another band gap with different topological invariants [Fig. 2b2].

The above two topological phases and associated phase transition can be described by the effective Hamiltonian of AA-stacking bilayer PC slabs with broken mirror and inversion symmetry [see detailed derivation in Supplementary Section C]:

$$\hat{H} = f_D + v_D(\delta k_x \hat{s}_0 \hat{\sigma}_x + \delta k_y \hat{s}_0 \hat{\sigma}_y) + \eta\Delta \hat{s}_z \hat{\sigma}_z + \eta\Sigma \hat{s}_0 \hat{\sigma}_z + w\hat{s}_x \hat{\sigma}_0, \quad (1)$$

where $\delta \vec{k} = (\delta k_x, \delta k_y)$ measures from the K point. $\hat{\sigma}_i$ and $\hat{s}_i$ are the Pauli matrices acting on sub-lattice and layer spaces, respectively. In equation (1), the second and third terms give two Dirac cones with the velocity $v_D$, the fourth (fifth) term describes the band gap opening under broken mirror symmetry (broken inversion symmetry) with a gap opening coefficient $\eta$, and the last term shows the interlayer coupling strength of $w$. The effective Hamiltonian implies that the phase transition boundary is given by [35]:

$$\Sigma^2 = \Delta^2 + (w/\eta)^2, \quad (2)$$

where the values of $w$ and $\eta$ can be numerically fitted from the frequency spectra in Fig. 2a. According to Eq. (2), the red curve in Fig. 3a plots the analytical phase transition boundary. The analytical results are in good agreement with those calculated by numerical simulation [red dots in Fig. 3a], which in turn proving the validity of the effective Hamiltonian. According to the effective Hamiltonian presented in Eq. (1), the layer-polarized PTI is characterized by the topological invariants of $C_L = 1$ and $C_K = 0$, and the layer-mixed PTI by $C_L = 0$ and $C_K = 1$ [35].

**Layer dependent edge states in layer-polarized photonic topological insulators**

The nonzero topological invariants indicate the nontrivial topology, implying the protected edge states at the domain wall between two topologically distinct PTIs. We first consider the topological domain wall which consists of the layer-polarized PTI with $C_L = 1$ (e.g., $s_{A1} = 130$ nm and $s_{A2} = 328.8$ nm with corresponding $\Delta = -91200$ nm$^2$ and $\Sigma = 0$) below the boundary and another layer-polarized PTI with $C_L = -1$ (e.g., $s_{A1} = 328.8$ nm and $s_{A2} = 130$ nm with corresponding $\Delta = 91200$ nm$^2$ and $\Sigma = 0$) above the boundary [Fig. 4a]. Its edge state dispersion is shown in Fig.

4b in which the light cone is shaded in grey. Edge states with the energy localizing near the boundary is marked by big circles while the bulk states are by small circles. Near the K valley, edge states with positive group velocities have fields localizing at layer1, while edge states with negative group velocities have fields localizing at layer2 [left and right panels of Fig. 4c]. Layer-direction locking edge states are observed. One feature of layer-direction locking edge states is the layer dependent transmission. As shown in Fig. 4d, when the source is put at layer1 at the left end, it excites the rightward light flow. The excited edge states are localized at layer1 [Fig. 4d]. On the other hand, when the source is put at layer2, it also excites the rightward light flow, but the excited edge states are localized at the layer2 [Fig. 4e].

**Protected transmission in layer-mixed photonic topological insulators**

Another kind of topological domain walls is formed by the layer-mixed PTI with $C_K = 1$ (e.g., $s_{A1} = 130$ nm and $s_{A2} = 130$ nm with corresponding $\Delta = 0$ and $\Sigma = -91200$ nm$^2$) below the boundary and the layer-mixed PTI with $C_K = -1$ (e.g., $s_{A1} = 328.8$ nm and $s_{A2} = 328.8$ nm with corresponding $\Delta = 0$ and $\Sigma = 91200$ nm$^2$) above the boundary [upper schematic in Fig. 5a]. Its edge state dispersion is shown in Fig. 5b. As the difference of $C_K$ is 2 crossing the domain wall, two edge states with positive group velocities are found near the K valley. Figure 5c illustrates the $|H_z|$ fields of two edge states at $k_x = -2\pi/3a$, showing the layer-mixed field distributions. Transmission of edge states at this domain wall is presented in Figs. 5d. $|H_z|$ varies alternatively in layer1 and layer2 because of the interference between two excited edge states. The oscillation length is determined by the momentum difference between two edge states. Such oscillation can be used to protect the high transmission even when a large defect is introduced. For example, a defected photonic waveguide with

5×6 unit cells removing at layer1 is considered [bottom panel of Fig. 5a]. Figure 5e shows that although edge states are blocked at layer1, the edge states can still pass along the additional channel at layer2, and the oscillation happens again after the defected cavity. High transmission of the bilayer domain wall is in a strong contrast to the low transmission of the monolayer domain wall [see in Supplementary Section D], showing the potential in protected light transport by using the additional layer pseudospin.

**Conclusion and Outlook**

In conclusion, we report layered PTIs by introducing the layer pseudospin into all-dielectric bilayer PC slabs. Layer-polarized and layer-mixed PTIs are observed, and the phase transition between them is demonstrated by a complete phase diagram. These are well illustrated by a model Hamiltonian by considering the nonzero interlayer coupling and symmetry breaking perturbations. The layer-direction locking behavior and high transmission are demonstrated. By employing the additional layer pseudospin, our work has demonstrated the band dispersion engineering capacity and light flow control in layered PTIs. The introduced all-dielectric bilayer PC slabs will become a powerful platform for emulating other topological states and studying other interesting physical phenomena.

**Acknowledgements**

This work was supported by National Natural Science Foundation of China (Grant Nos. 11704422, 61775243, 11522437, 61471401, and 11761161002), Natural Science Foundation of Guangdong Province (Grant No. 2018B030308005), and Science and Technology Program of Guangzhou (Grant No. 201804020029).

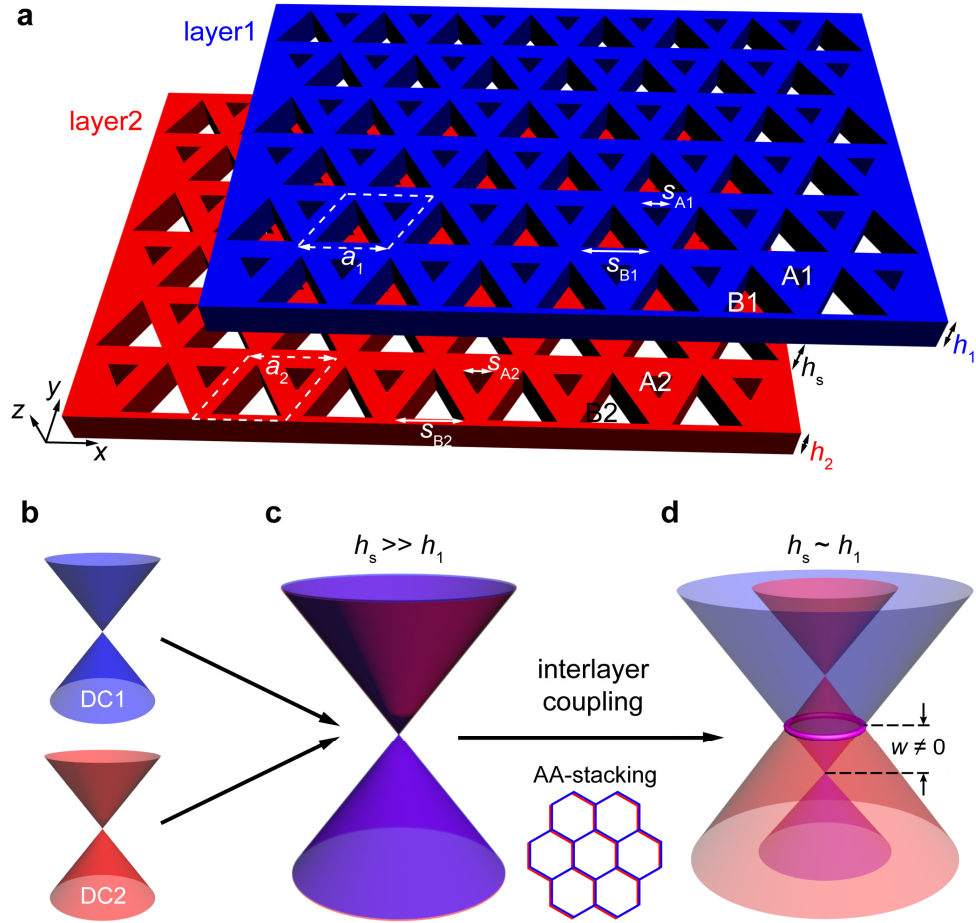

**Figure 1 | Bilayer photonic crystal (PC) slabs. a**, Schematic of an all-dielectric bilayer PC slab with the upper layer (layer1) shifting to show the lower layer (layer2). Each PC slab consists of a dielectric membrane ($\varepsilon = 12$) with a honeycomb lattice of triangular air-holes (i.e., A1 and B1 in layer1 while A2 and B2 in layer2). These two PC slabs are separated by a homogeneous spacer. Structural parameters: the lattice constant of PC slabs ($a_1$ and $a_2$), the thickness of two layers ($h_1$ and $h_2$), the thickness of spacer ($h_s$), the lateral size of triangular air-holes ($s_{A1}$ & $s_{B1}$, and $s_{A2}$ & $s_{B2}$). Throughout this work, the subscripts of "1", "2" and "s" represent layer1, layer2 and spacer, respectively. **b-d**, Schematics of the evolution of double Dirac cones under the interlayer coupling. **b**, Transverse-electric like states of each slab form a gapless Dirac cone (DC) around the corner of Brillouin zone. **c**, Doubly degenerate Dirac cones occur when two slabs are far apart, i.e., $h_s \gg h_1$. **d**, When the interlayer coupling is turned on (i.e., $w \neq 0$) for AA-stacking bilayer PC slabs, two Dirac cones are shifted oppositely along the frequency axis. A frequency nodal ring occurs around the K point (marked in pink), serving as an ideal starting point to have PTIs.

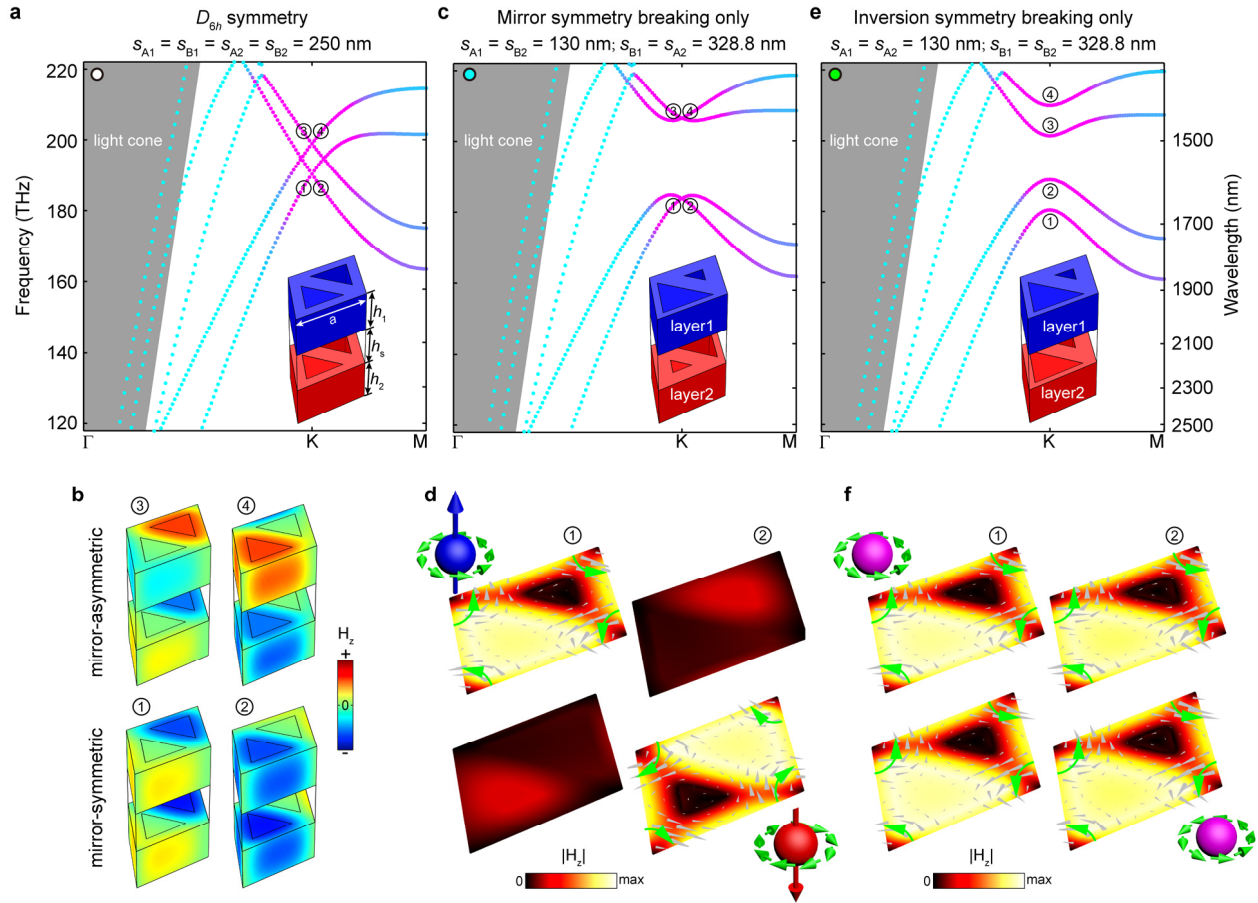

**Figure 2 | Layered photonic topological insulators (PTIs). a-f**, Bulk band structures and eigen-fields of AA-stacking bilayer PC slabs without (**a**, **b**) and with (**c-f**) perturbations. The lattice constants are $a = a_1 = a_2 = 460$ nm, and the thicknesses of slabs and spacer are $h_1 = h_2 = h_s = 220$ nm. The light cone is shaded in grey. The schematic of unit cell is shown in the inset. **a**, The $D_{6h}$ symmetric PC slab with equal air-hole sizes of $s_{A1} = s_{B1} = s_{A2} = s_{B2} = 250$ nm has two frequency shifted Dirac cones and a frequency nodal ring enclosing the K point. **b**, Eigen fields ($H_z$) for four bulk states at the K point. Two mirror-symmetric states (① and ② patterns) appear at the lower frequency, while two mirror-asymmetric states (③ and ④ patterns) show at the higher frequency. **c**, The mirror symmetry breaking bilayer PC slab with $s_{A1} = s_{B2} = 130$ nm and $s_{B1} = s_{A2} = 328.8$ nm has a directional band gap and two pairs of degenerate states at the K point. **d**, The absolute value of $H_z$ (colors) and the power flux (grey arrows) at the central plane of layer1 and layer2 for two lowest bulk states at the K point. The two-fold degeneracy is between the pseudospin-up anticlockwise state and pseudospin-down clockwise state (3D schematics), confirming the photonic analogue of spin-orbit coupling when the 'layer' is treated as a pseudospin. **e**, The inversion symmetry breaking bilayer PC slab with $s_{A1} = s_{A2} = 130$ nm and $s_{B1} = s_{B2} = 328.8$ nm has a directional band gap. **f**, The absolute value of $H_z$ (colors) and the power flux (grey arrows) at the central plane of layer1 and layer2 for two lowest bulk states at the K point. These two states belong to the same orbital states (3D schematics). Note that the green arrows in **d** and **f** are used to guide eyes.

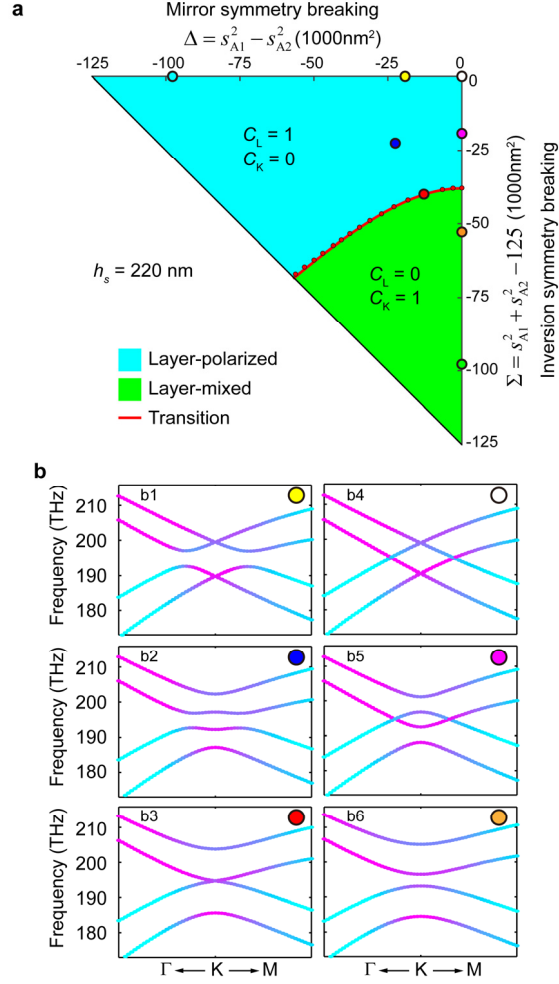

**Figure 3 | Phase diagram of layered PTIs. a**, Phase diagram of AA-stacking bilayer PC slabs with the same spacer thickness of $h_s = 220$ nm but varied $s_{A1}$ and $s_{A2}$. The amplitude of mirror (inversion) symmetry breaking is given by $\Delta = s_{A1}^2 - s_{A2}^2$ ($\Sigma = s_{A1}^2 + s_{A2}^2 - 125000\,\text{nm}^2$). The layer-polarized (layer-mixed) PTIs are characterized by $C_L = 1$ and $C_K = 0$ ($C_L = 0$ and $C_K = 1$) and they are shaded in cyan (green). The phase transition boundary between two topologically distinct PTIs is marked in red (small red dots for numerical results and the red curve for analytical results). Eight representative bilayer PC slabs presented in (**b**) and Fig. 2 are labeled by big dots. **b**, Bulk band strucutres of six representative bilayer PC slabs, showing the phase transition. Structural parameters: (b1) $\Delta = 0$, $\Sigma = 0$, (b2) $\Delta = 0$, $\Sigma = -19200$ nm$^2$, (b3) $\Delta = 0$, $\Sigma = -52800$ nm$^2$, (b4) $\Delta = -19200$ nm$^2$, $\Sigma = 0$, (b5) $\Delta = -22500$ nm$^2$, $\Sigma = -22500$ nm$^2$, (b6) $\Delta = -39970$ nm$^2$, $\Sigma = -12829$ nm$^2$. The color of up-right dot in each subfigure is in accordance with that in (**a**).

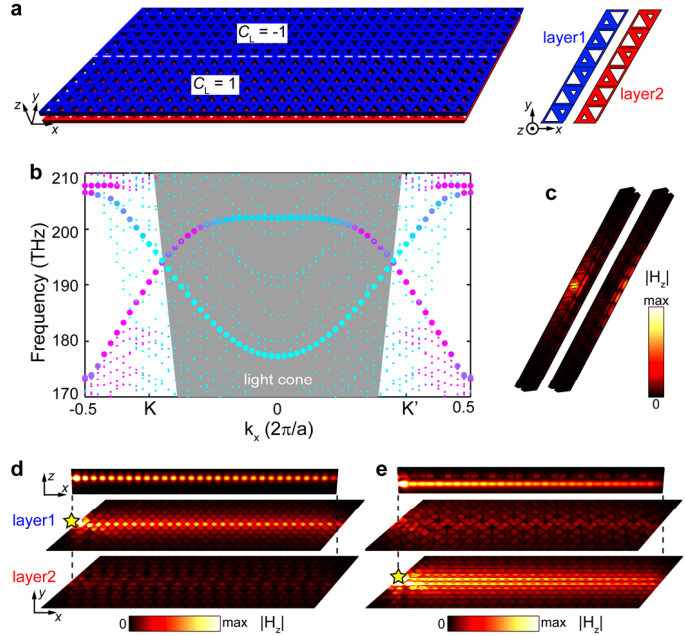

**Figure 4 | Layer dependent edge states and selective light refraction. a**, The schematics of the topological domain wall between two layer-polarized PTIs. Right insets: The 2D views of layer1 and layer2 near the boundary. **b**, Band dispersions of eigen states. Big circles mark the edge states with fields localizing at the boundary, while small circles mark the bulk states. Near the K valley, edge states with positive (negative) group velocities are localized at layer1 (layer2). Edge states near the K' valley can be obtained by considering the time-reversal symmetry. The light cone is shaded in grey. **c**, Representative fields ($|H_z|$) for edge states at $k_x = -2\pi/3a$ with fields locating at layer1 (left) and at layer2 (right). **d** & **e**, Layer dependent transmission when the incident source (yellow star) is put at the left end of (**d**) layer1, and (**e**) layer2.

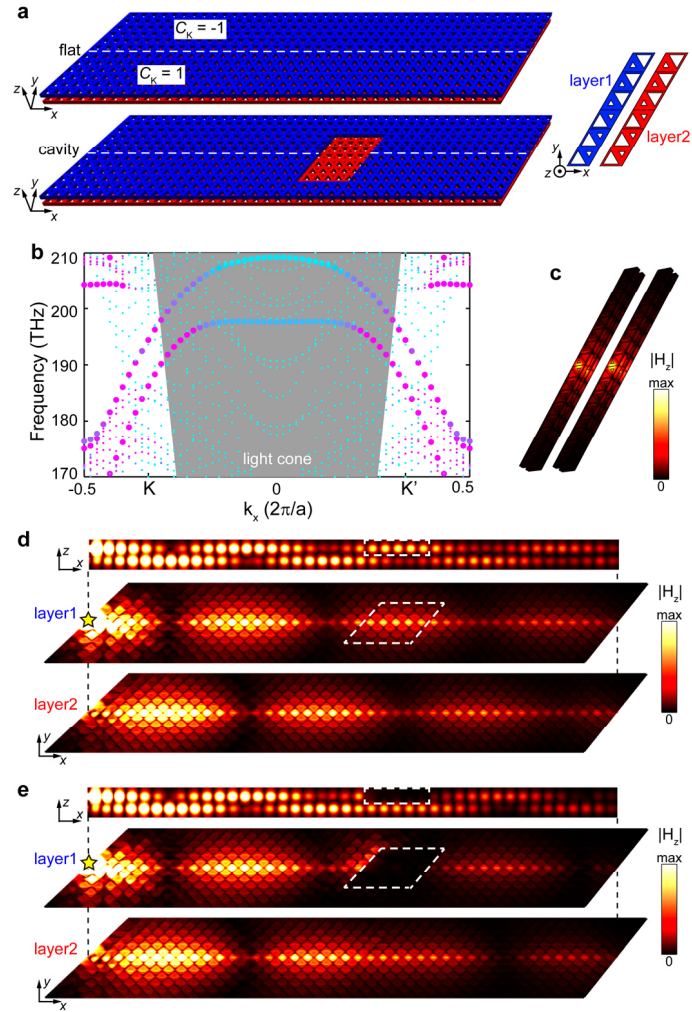

**Figure 5 | Protected transmission at the topological domain wall between two layer-mixed PTIs.**
**a**, The schematics of topological domain wall between two layer-mixed PTIs without a defect (up) and with a large cavity (down). Right insets: The 2D views of layer1 and layer2 near the boundary. **b**, Band dispersions of eigen states. At the K (K') valley, both edge states have positive (negative) group velocities, with fields mixing at both layers. Big circles mark the edge states with fields localizing at the boundary, while small circles mark the bulk states. The light cone is shaded in grey. **c**, Representative fields ($|H_z|$) for two edge states at $k_x = -2\pi/3a$, showing the layer-mixed field distributions. **d**, Transmitted electromagnetic fields ($|H_z|$) for the flat waveguide. The oscillation of field amplitude confirms the layer-mixed property. **e**, Transmitted electromagnetic fields ($|H_z|$) for the waveguide with a cavity. High transmission is preserved. In (**d**) and (**e**), sources (yellow star) are excited at layer1, and the boundaries of the cavity are outlined by white dash polygons.